\documentclass{article}

\usepackage{arxiv}

\usepackage[utf8]{inputenc} 
\usepackage[T1]{fontenc}    
\usepackage{hyperref}       
\usepackage{url}            
\usepackage{booktabs}       
\usepackage{amsfonts}       
\usepackage{nicefrac}       
\usepackage{microtype}      
\usepackage{lipsum}		
\usepackage{graphicx}
\usepackage{natbib}
\usepackage{doi}

\hypersetup{
  colorlinks   = true, 
  urlcolor     = blue, 
  linkcolor    = red, 
  citecolor   = green 
}
\setcitestyle{numbers,open={[},close={]}}
\usepackage{subcaption}
\usepackage{adjustbox}
\usepackage{multirow}

\title{Diffusion Probabilistic Models beat GAN on Medical 2D Images}

\author{
        \href{https://orcid.org/0000-0002-7413-2570}{\includegraphics[scale=0.06]{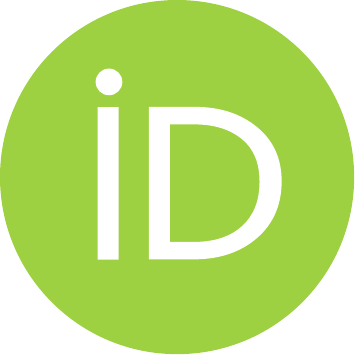}\hspace{1mm}Gustav Müller-Franzes} 
	\And
        \href{https://orcid.org/0000-0003-4016-9129}{\includegraphics[scale=0.06]{orcid.pdf}\hspace{1mm}Jan Moritz Niehues} 
	\And
        \href{https://orcid.org/0000-0001-5089-3589}{\includegraphics[scale=0.06]{orcid.pdf}\hspace{1mm}Firas Khader} 
	\And
         \href{https://orcid.org/0000-0003-1015-7733}{\includegraphics[scale=0.06]{orcid.pdf}\hspace{1mm}Soroosh Tayebi Arasteh} 
	\And
         \href{https://orcid.org/0000-0002-2405-640X}{\includegraphics[scale=0.06]{orcid.pdf}\hspace{1mm}Christoph Haarburger} 
	\And
         \href{https://orcid.org/0000-0001-8696-2363}{\includegraphics[scale=0.06]{orcid.pdf}\hspace{1mm}Christiane Kuhl} 
	\And
         \href{https://orcid.org/0000-0002-6783-649X}{\includegraphics[scale=0.06]{orcid.pdf}\hspace{1mm}Tianci Wang} 
	\And
         \href{https://orcid.org/0000-0002-8636-6462}{\includegraphics[scale=0.06]{orcid.pdf}\hspace{1mm}Tianyu Han} 
         \And
         \href{https://orcid.org/0000-0002-5267-9962}{\includegraphics[scale=0.06]{orcid.pdf}\hspace{1mm}Sven Nebelung} 
	\And
	\href{https://orcid.org/0000-0002-3730-5348}{\includegraphics[scale=0.06]{orcid.pdf}\hspace{1mm}Jakob Nikolas Kather} 
        \thanks{Contributed eqally}
        \And
	\href{https://orcid.org/0000-0002-9605-0728}{\includegraphics[scale=0.06]{orcid.pdf}\hspace{1mm}Daniel Truhn} 
        \footnotemark[1]
}

\date{}

\renewcommand{\undertitle}{}

\hypersetup{
pdftitle={A template for the arxiv style},
pdfsubject={cs.CV},
pdfauthor={},
pdfkeywords={},
}

\begin{document}
\maketitle

\begin{abstract}
The success of Deep Learning applications critically depends on the quality and scale of the underlying training data. 
Generative adversarial networks (GANs) can generate arbitrary large datasets, but diversity and fidelity are limited, which has recently been addressed by denoising diffusion probabilistic models (DDPMs) whose superiority has been demonstrated on natural images.  
In this study, we propose Medfusion, a conditional latent DDPM for medical images. We compare our DDPM-based model against GAN-based models, which constitute the current state-of-the-art in the medical domain.
Medfusion was trained and compared with (i) StyleGan-3 on n=101,442 images from the AIROGS challenge dataset to generate fundoscopies with and without glaucoma, (ii) ProGAN on n=191,027 from the CheXpert dataset to generate radiographs with and without cardiomegaly and (iii) wGAN on n=19,557 images from the CRCMS dataset to generate histopathological images with and without microsatellite stability.
In the AIROGS, CRMCS, and CheXpert datasets, Medfusion achieved lower (=better) FID than the GANs (11.63 versus 20.43, 30.03 versus 49.26, and 17.28 versus 84.31). Also, fidelity (precision) and diversity (recall) were higher (=better) for Medfusion in all three datasets.
Our study shows that DDPM are a superior alternative to GANs for image synthesis in the medical domain.

\end{abstract}


\section{Introduction}
The performance of deep learning crucially depends on the size of the available training set \cite{cho2016, samala2019}. However, in the medical domain, data is often not publicly available and large data pools cannot be sourced from multiple sites because of privacy issues. In the past, generative adversarial models (GANs) have been used to address these problems \cite{wang2021b}. Generative models have a variety of possible applications, from sharing data and circumventing legal or ethical difficulties \cite{han2020c} to reducing the need for data through modality translation \cite{armanious2020} and improving deep learning performance \cite{han2020c, krause2021}. However, generating meaningful medical data is hard, since medical diagnosis often depends on subtle changes in the appearance of complex organs and it is often more challenging than image classification on natural images. In addition, GANs suffer from inherent architectural problems such as the failure to capture true diversity, mode collapse, or unstable training behavior \cite{saxena2020}. Thus, particular emphasis needs to be put on the generation of high-quality synthetic medical data.
Recently, denoising diffusion probabilistic models (DDPMs) \cite{ho2020} and latent DDPMs \cite{rombach2022} have shown state-of-the-art results and were able to outperform GANs on natural images \cite{dhariwal2021}. While DDPMs have already demonstrated their superiority over GANs on natural images, a wide-scale direct comparison of latent DDPMs to GANs on medical images covering multiple domains has so far not been done.
We found two studies that directly compared DDPMs and GANs for medical image synthesis in specific use cases. Pinaya et al. \cite{pinaya2022} used a latent DDPM to generate 3D brain MRI images, which was trained and conditioned on 31,740 T1-weighted images from the UK Biobank with covariables such as age and sex. They found that their latent DDPM outperforms LSGAN and VAE-GAN. In a similar study, a DDPM has been used to generate 3D brain MRI images and was trained but not conditioned on 1500 T1-weighted images from the ICTS dataset \cite{dorjsembe2022}. In a quantitative comparison, the DDPM could outperform a 3D-$\alpha$-WGAN but not a CCE-GAN. However, when two radiologists with 15 years of experience were asked to classify the images as real or fake, 60\% of the DDPM-generated images were rated as real but none of the GAN images.
These studies show that (latent) DDPMs are a promising alternative to GANs also in the medical domain. However, tests were limited to MRI and the brain and focused on 3D image generation.
In this study, we propose Medfusion, a conditional latent DDPM for medical images. We compare our DDPM-based model against GAN-based models by using images sourced from ophthalmology, radiology and histopathology and demonstrate that DDPM beat GANs in all relevant metrics. To foster future research, we make our model publicly available as open-source to the scientific community.

\section{Material and Methods}

\subsection*{Ethics statement}
All experiments were conducted in accordance with the Declaration of Helsinki and the International Ethical Guidelines for Biomedical Research Involving Human Subjects by the Council for International Organizations of Medical Sciences (CIOMS). The study has additionally been approved by the local ethical committee (EK 22-319).

\subsection*{Datasets}
In this retrospective study, three publicly available datasets were used. 
First, the AIROGS \cite{zotero-1190} challenge train dataset, containing 101,442 256x256 RGB eye fundus images from about 60,357 subjects of which 98,172 had “no referable glaucoma” and 3,270 with “referable glaucoma”. Sex and age of the subjects were unknown. 
Second, the CRCDX \cite{katherjakobnikolas2020} dataset, containing 19,958 color-normalized 512x512 RGB histology colorectal cancer images at a resolution of 0.5 µm/px. Half of the images were microsatellite stable and microsatellite instable, respectively. Sex and age of the subjects were unknown. 
Third, the CheXpert \cite{irvin2019} train dataset, containing 223,414 gray-scaled chest radiographs of 64,540 patients. Images taken in lateral position were excluded, leaving 191,027 images from 64,534 patients. All images were scaled to 256x256 and normalized between -1 and 1, following the pre-processing routine in \cite{han2020c}. Of the remaining radiographs,  23,385 showed an enlarged heart (cardiomegaly), 7,869 showed no cardiomegaly, and 159,773 had an unknown status. Labels with unknown status were relabeled as in \cite{han2020c} such that 160,935 images had no cardiomegaly and 30,092 had cardiomegaly, respectively. Mean age of the 28,729 female and 35,811 male patients was 60±18 years at examination.

\subsection*{Model architecture and training details }
Two types of generative models were used in this study.

First, classical generative adversarial networks (GANs) as introduced by Goodfellow et al. \cite{goodfellow2014}.  We aimed to use GANs that exhibited state-of-the-art quality on the respective datasets to allow a fair comparison with the diffusion model. 
For the CheXpert dataset, we used a pre-trained progressive growing GAN (proGAN) from \cite{han2020c}. Chest x-rays generated by this GAN were barely differentiably by three inexperienced and three experienced radiologists \cite{han2020c}. Furthermore, this GAN architecture has already been used for data augmentation and has led to higher downstream performance as compared to traditional augmentation \cite{sundaram2021}. 
For the CRCDX dataset, we employed a pre-trained, conditional GAN (cGAN) as described in \cite{krause2021}. This GAN has been shown to produce realistic-looking histological cancer images in a blinded test with five readers and the authors were able to show that a classifier benefits from using generated images during training. 
No pre-trained GAN was available for the AIROGS dataset at the time of writing. Therefore, StyleGAN-3 \cite{karras2021} was used as it incorporates most of the latest GAN developments and has shown state-of-the-art results. We used the default settings that were proposed by the authors for images of 256x256 pixels and trained the model for 3000 iterations.

\begin{figure}
    \centering
    \includegraphics[width=\textwidth]{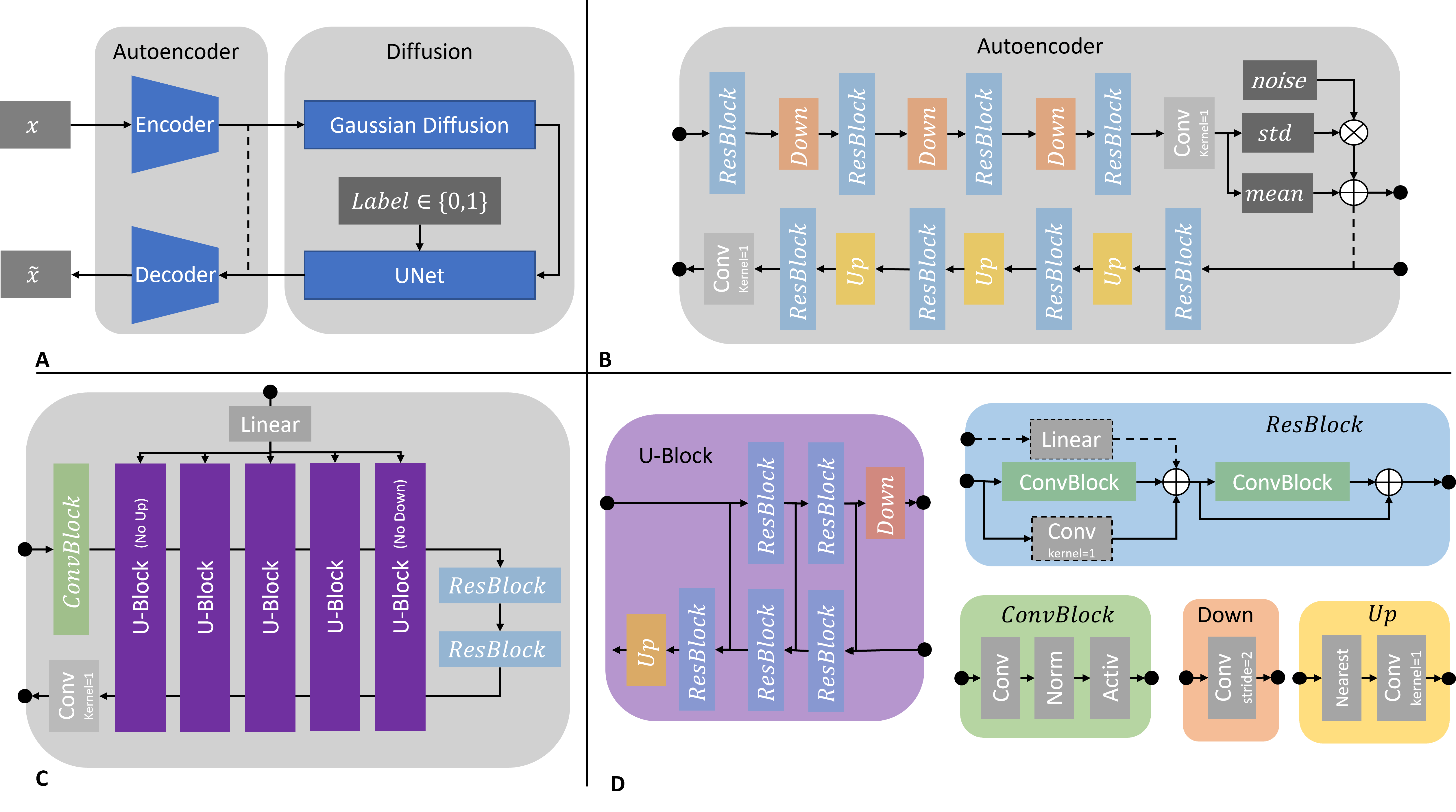}
    \caption{Illustration of the Medfusion model. A: General overview of the architecture. B:  Details of the autoencoder with a sampling of the latent space via the reparameterization trick at the end of the encoder and direct connection (dashed lines) into the decoder (only active for training the autoencoder). C: Detailed view of the UNet with a linear layer for time and label embedding. D: Detailed view of the submodules. If not specified otherwise a convolution kernel size of 3x3, GroupNorm with 8 groups, and Swish activation was used.   }
    \label{fig:fig1}
\end{figure}

Second, our proposed Medfusion model (Figure \ref{fig:fig1}) is based on the Stable Diffusion model \cite{rombach2022}. It consisted of two parts:  An Autoencoder that encoded the image space into a compressed latent space and a DDPM \cite{ho2020}.  Both parts were trained in two subsequent phases. In the first training phase, the Autoencoder was trained to encode the image space into an 8-times compressed latent space of size 32x32 and 64x64 for the 256x256 and 512x512 input space, respectively. During the training phase, the latent space was directly decoded back into image space and supervised by a multi-resolution loss function, which was described in the supplemental material. In the second training phase, the pre-trained Autoencoder encoded the image space into a latent space, which was then diffused into Gaussian noise using t=1000 steps. A UNet \cite{ronneberger2015} model was used to denoise the latent space. The weights of the Autoencoder were frozen during the second-training phase. Samples were generated with a Denoising Diffusion Implicit Model (DDIM) \cite{song2022} and t=150 steps. We motivate our choice of steps in the supplemental material.

\subsection*{Experimental design}
The study was divided into two sub-studies:
First, we investigated whether the capacity of the autoencoder in the Medfusion model was sufficient to encode the images into a latent, highly compressed space and decode the latent space back into the image space without losing relevant medical details. It was also investigated whether the autoencoder of the Stable Diffusion Model (pre-trained on natural images) could be used directly for medical images, i.e. without further training on medical images and loss of medically relevant image details.
Second, we compared the images generated by Medfusion and the GANs quantitatively and qualitatively.  For the quantitative evaluation, we would like to refer to the statistics section in which we go into detail about the individual metrics. For the qualitative assessment, real, GAN-generated, and Medfusion-generated images were compared side-by-side.

\subsection*{Statistical analysis}
All statistical analyses were performed using Python and implemented in TorchMetrics \cite{detlefsen2022}.  To compare sample quality across models, the following metrics were used.
First, the Fréchet Inception Distance (FID) \cite{NIPS2017_8a1d6947}, that has become a standard metric for quality comparisons of generative models \cite{borji2021} and measures the agreement of the real and synthetic images by comparing the features of the deepest layer of the Inception-v3 \cite{szegedy2016} model. 
Second, the Improved Precision and Recall \cite{NIPS2016_8a3363ab} metric, that measures the fidelity as the overlap of synthetic and real features relative to the entire set of synthetic features (precision) and the diversity as the overlap relative to the entire set of real features (recall). Following a previous study \cite{dhariwal2021}, Inception-v3 was used instead of the original proposed VGG-16 \cite{simonyan2015} to extract features. 
Third, the Multiscale Structural Similarity Index Measure (MS-SSIM) \cite{wang2003} that is a generalized version of the SSIM \cite{wang2004} by applying SSIM at different resolutions. The SSIM measures image distortion by the structural information change that is expressed by comparing luminance, contrast, and structure. 
To ensure consistency between model comparisons, a reference batch was used for the AIROGS, CRCDX, and CheXpert dataset with 6540, 19,958, and 15,738 equally distributed images of both classes. Of note, the used metrics depend on the reference subset and implementation \cite{parmar2022} and are not directly comparable with other studies.

\subsection*{Implementation and data availability}
All experiments were implemented in Python v3.8 and were executed on a computer with an Nvidia RTX 3090. The datasets can be downloaded directly from the websites of the referenced authors.  Source code for the StyleGan-3, ProGAN, and cGAN are available at \url{https://github.com/NVlabs/stylegan3}, \url{https://github.com/peterhan91/Thorax_GAN} and \url{https://github.com/mjendrusch/pytorch-histogan}. Source code for the Medfusion Model is available at \url{https://github.com/mueller-franzes/medfusion} under an open-source license.

\section{Results}

\subsection*{High reconstruction capacity of Medfusion’s Autoencoder}
Sicne the quality of the diffusion-generated images is limited by the reconstruction quality of the autoencoder, we first investigated possible image quality losses due to the autoencoder architecture. To evaluate the maximum possible quality, samples in the reference batches were encoded and decoded by the Autoencoder. Subsequently, the MS-SSIM and mean squared error (MSE) between the input images and reconstructed images were calculated and averaged (Table \ref{tab:table1}). Both metrics indicated a nearly perfect (MS-SSIM = 1, MSE=0) reconstruction of the images in the AIROGS and CheXpert dataset. Reconstruction quality in the CRCDX dataset was good but lower compared to AIROGS or CheXpert datasets, most likely due to the four times higher resolution. Since these metrics were measured on the reference set which is part of the training set, these values can be considered as an upper bound (for MS-SSIM) and lower bound (for MSE), respectively. The results on the publicly available test set of the CheXpert and CRCDX dataset were however nearly identical to the results from the reference set and are available in the supplemental materials.
This experiment demonstrates that the autoencoder architecture of the DDPM does not restrict image quality of synthesized in terms of numeric metrics.

\begin{table}[]
        \caption{Medfusion Autoencoder reconstruction quality. Values represent mean±standard deviation, MSE = Mean Squared Error, MS-SSIM = Multiscale Structural Similarity Index Measure}
        \centering
        \begin{tabular}{@{}cccc@{}}
        \toprule
        \multicolumn{1}{l}{} & AIROGS      & CRCDX       & CheXpert    \\ \midrule
        MS-SSIM              & 0.981±0.007 & 0.901±0.040 & 0.994±0.001 \\
        MSE ($10^{-5}$)         & 11±7         & 541±305     & 25±9      \\ \bottomrule
        \end{tabular}
        \label{tab:table1}
\end{table}

\subsubsection*{Dataset-specific reconstruction challenges}
To investigate, if there are qualitative autoencoder reconstruction errors in the autoencoding process, we compared the original and reconstructed images side-by-side. This confirmed the numerically measured high reconstruction quality but revealed dataset-specific reconstruction errors (Figure \ref{fig:fig2}). The compression in the autoencoding stage resulted in subtle structural changes in the fundus images, color changes in the histology images, and a loss of sharpness in the thorax images. This demonstrates that DDPM could be further enhanced by making use of a better autoencoding architecture. To investigate, if this can be remedied with less compression during the autoencoding stage, we performed an additional experiment:

\begin{figure}
    \centering
    \includegraphics[width=\textwidth]{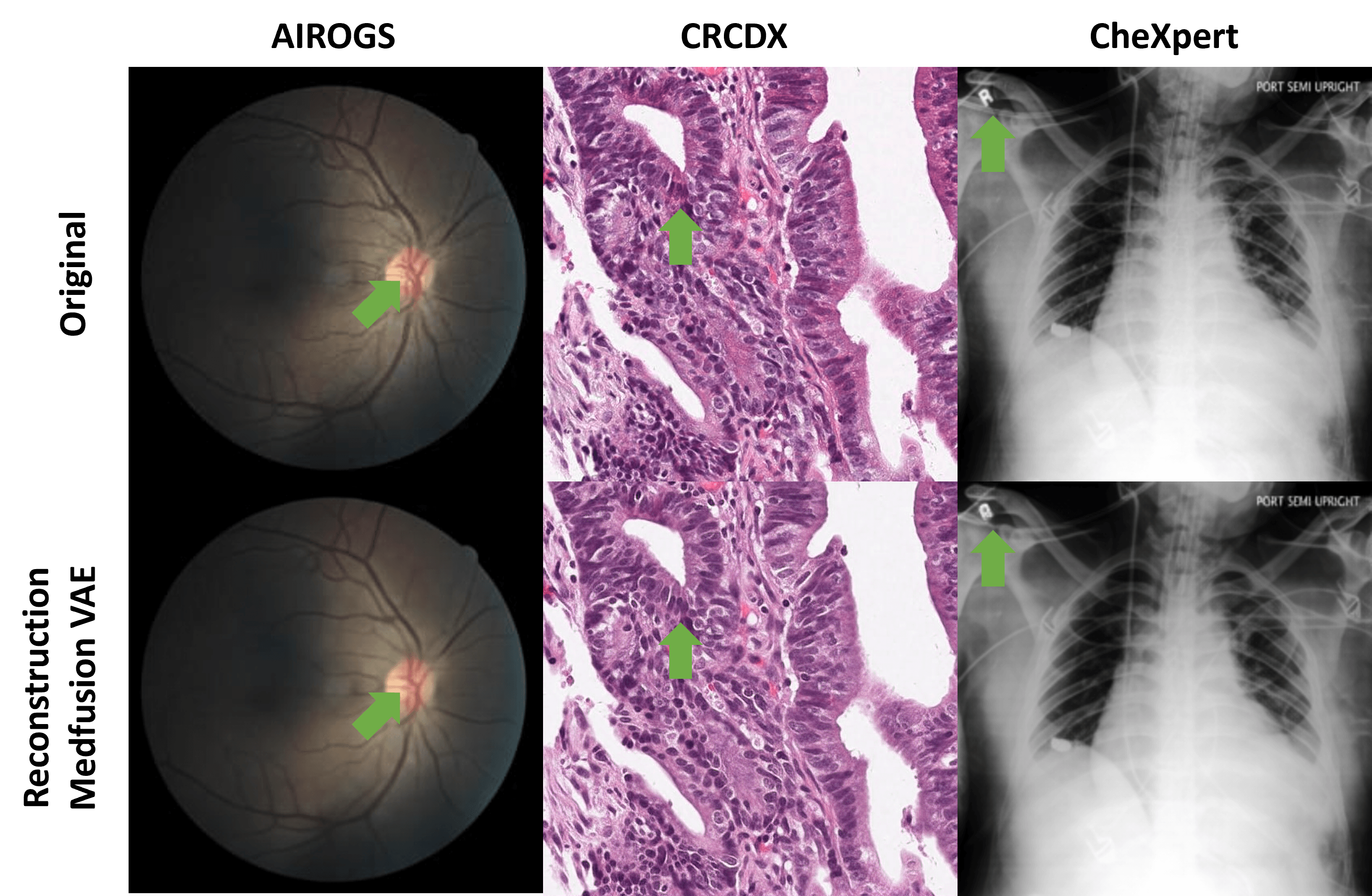}
    \caption{Reconstruction quality of Medfusion Variational Autoencoder (VAE).  Original images (first row) and reconstructed images (second row)  by the VAE in the AIRGOS, CRCDX, and CheXpert dataset.  In the eye fundus images, fine deviations from the original images were apparent in the veins of the optical disc (green arrow). Slight deviations in the color tone (green arrow) could be observed in the CRCDX dataset. In the CheXpert dataset, letters (green arrow) became blurry after reconstruction.  }
    \label{fig:fig2}
\end{figure}

\subsubsection*{Universal Autoencoders - 4 channels are not enough for medical images}
\begin{figure}
    \centering
    \includegraphics[width=\textwidth]{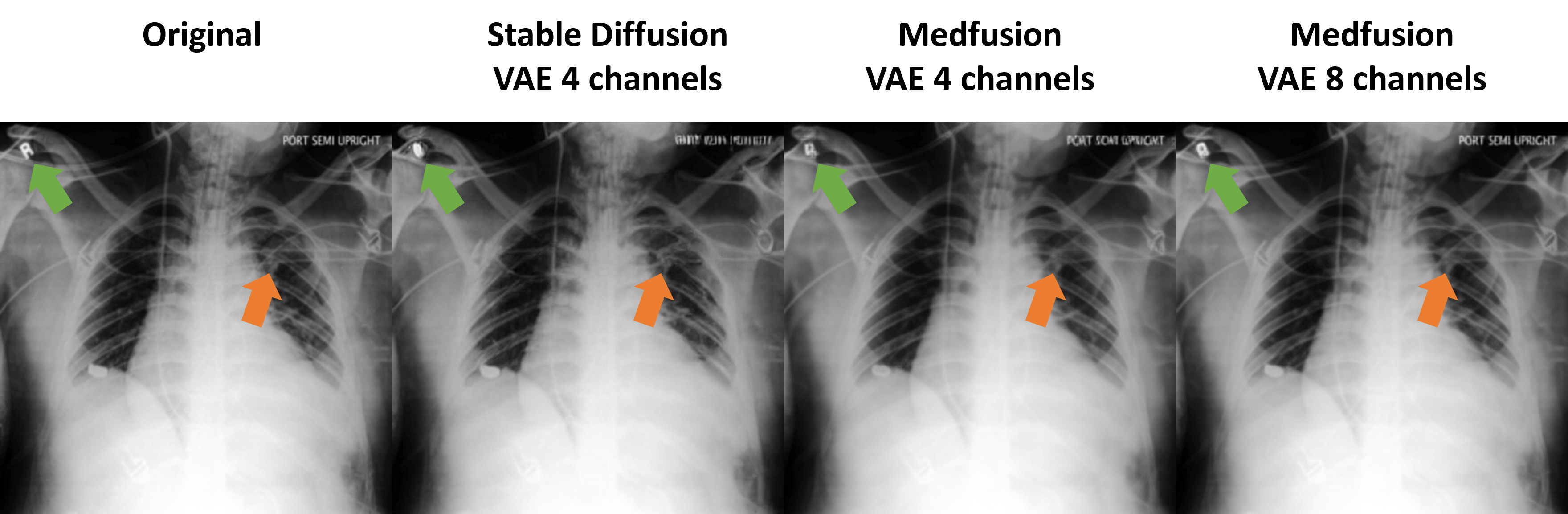}
    \caption{Reconstruction quality comparison. Both the out of the box VAE and the specifically trained VAE exhibit artifacts that may affect diagnostic accuracy. Here, lead cables are not reconstructed properly. Only when the number of channels is increased to eight are such small structures accurately reconstructed. }
    \label{fig:fig3}
\end{figure}

A comparison with the autoencoder taken out of the box from the Stable Diffusion Model demonstrated that the reconstruction of medical images work well with an autoencoder pre-trained on natural images (Table \ref{tab:table2}). However, when comparing images side-by-side, Stable Diffusion showed characteristic reconstruction errors in the CheXpert dataset when Stable Diffusion’s default VAE with 4 channels was used  (Figure \ref{fig:fig3}). Although less severe, reconstruction errors were also evident in Medfusion's VAE reconstructions. A further increase in the number of trainable parameters did not seem reasonable because Stable Diffusion 4-channel VAE already had about three times as many parameters as Medfusion’s 4-channel VAE (24 million). Therefore, we increased the number of channels from 4 to 8 instead, which resulted in a notable quality gain at the cost of compression ratio.
These results demonstrate that DDPM’s autoencoding architecture can benefit from a higher number of channels during the autoencoding stage.

\begin{table}[]
        \caption{Stable Diffusion Autoencoder reconstruction quality.Values represent mean±standard deviation, MSE = Mean Squared Error, MS-SSIM = Multiscale Structural Similarity Index Measure}
        \centering
        \begin{tabular}{@{}cccc@{}}
        \toprule
        \multicolumn{1}{l}{} & ARIOGS      & CRCDX       & CheXpert    \\ \midrule
        MS-SSIM              & 0.973±0.010 & 0.870±0.049 & 0.973±0.006 \\
        MSE ($10^{-5}$)           & 22±10       & 640±383     & 90±30       \\ \bottomrule
        \end{tabular}
        \label{tab:table2}
\end{table}

\subsection*{Medfusion outperforms GANs}
\begin{figure}
    \centering
    \includegraphics[width=\textwidth,height=\textheight,keepaspectratio]{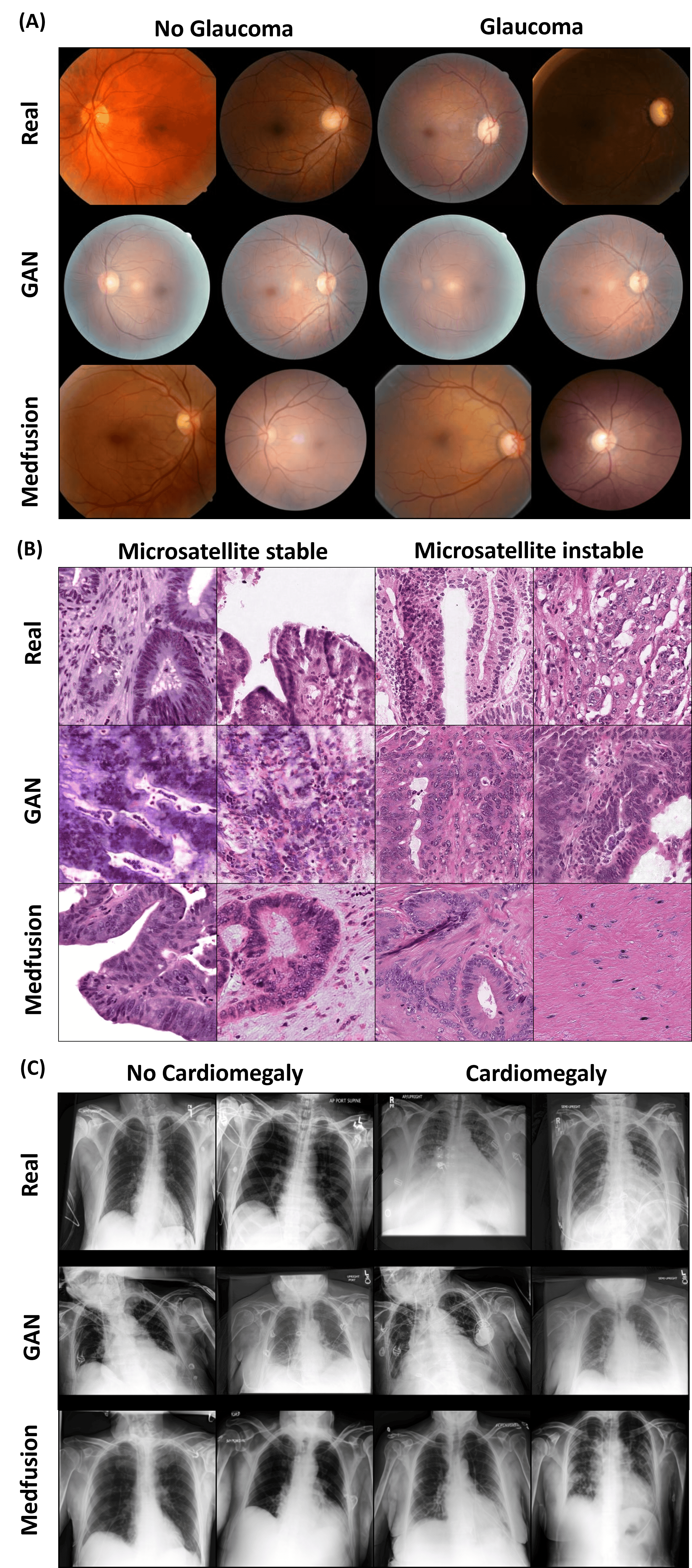}
    \caption{Qualitative image generation comparisons. Real images (first row), generated image by the GAN (second row) and Medfusion (third row), columns 1-2 conditioned on no glaucoma and columns 3-4  glaucoma (A), microsatellite stable and microsatellite instable (B) and no cardiomegaly and cardiomegaly (C).  }
    \label{fig:fig4}
\end{figure}

When comparing real and synthetic images based on the FID metric, we found that Medfusion generated more realistic-looking images in all three datasets than the corresponding GAN models (Table \ref{tab:table3}).  This was confirmed by the numeric metrics: Precision and Recall values showed that Medfusion had higher fidelity while it maintained greater diversity among the images as compared with the GAN models. Sample images for qualitative comparison are given in Figure \ref{fig:fig4}. We found that DDPM generated consistently more realistic and more diverse synthetic images than GANs. Together, these data show that DDPMs are superior to GANs both in terms of quantitative and qualitative metrics.

\begin{table}[]
\caption{Quantitative image generation comparisons. Models include Generative Adversarial Networks (StyleGan-3, cGAN, and ProGAN) and our proposed Medfusion model.  FID = Fréchet Inception Distance}
\centering
\begin{tabular}{@{}ccccc@{}}
\toprule
Dataset  & Model      & FID ↓ & Precision ↑ & Recall ↑ \\ \midrule
AIROGS   & StyleGan-3 & 20.43 & 0.43        & 0.19     \\
AIROGS   & Medfusion  & 11.63 & 0.70        & 0.40     \\
CRCDX    & cGAN       & 49.26 & 0.64        & 0.02     \\
CRCDX    & Medfusion  & 30.03 & 0.66        & 0.41     \\
CheXpert & ProGAN     & 84.31 & 0.30        & 0.17     \\
CheXpert & Medfusion  & 17.28 & 0.68        & 0.32     \\ \bottomrule
\end{tabular}
\label{tab:table3}
\end{table}

We provide a website with sample images to the scientific community so that a straightforward and more comprehensive review of Medfusion’s image quality is possible. The website can be accessed at: \url{https://huggingface.co/spaces/mueller-franzes/medfusion-app}. 

\subsection*{GAN Synthesized Images Exhibit Characteristic Artefacts}
\begin{figure}
    \centering
    \includegraphics[width=\textwidth]{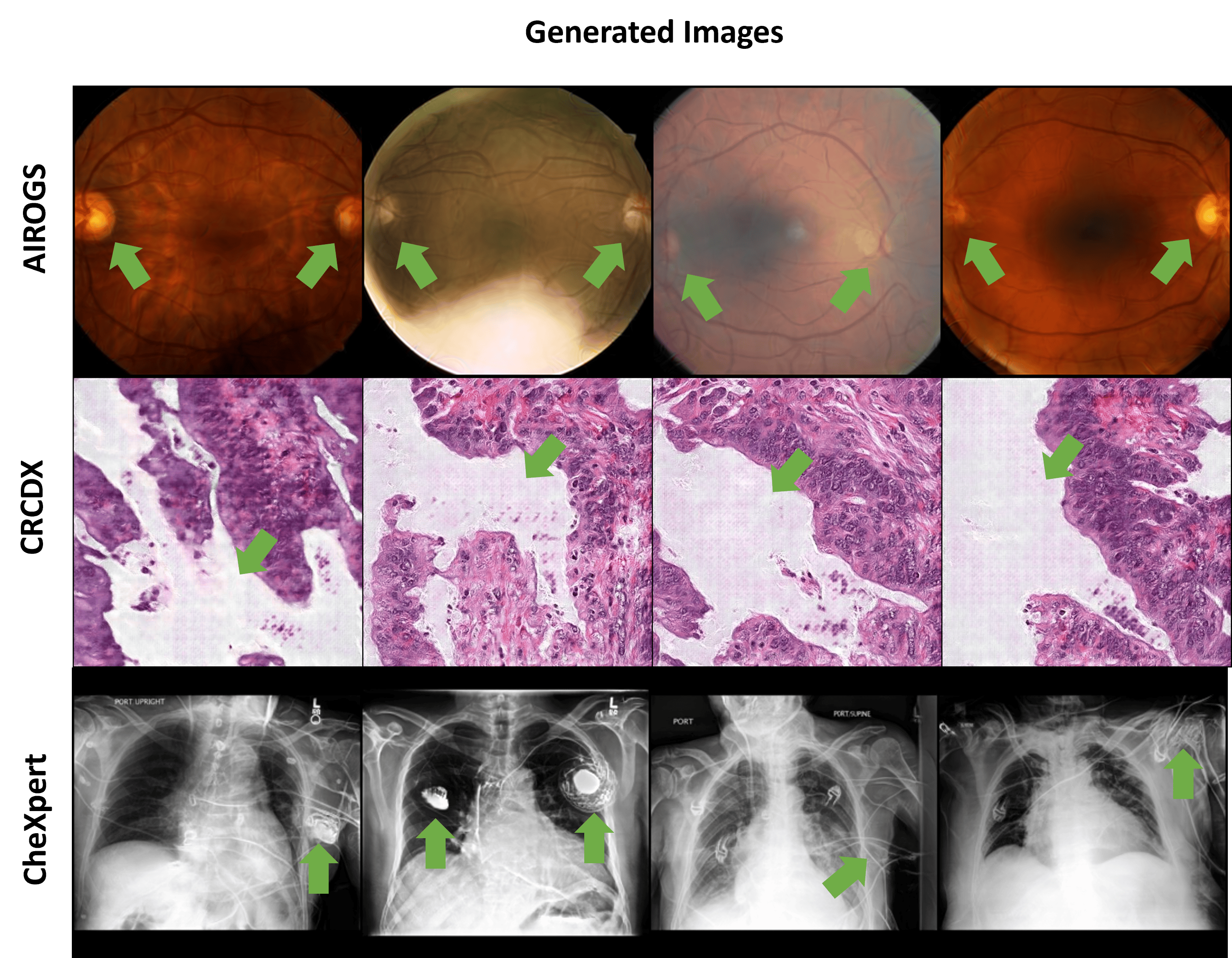}
    \caption{GAN-generated images that can be easily identified as synthetic. Synthetic images were easily identifiable because of two optical discs in eye fundus images, artificial grid patterns in histology images, and fuzzy borders and irregular appearances of medical devices in chest x-ray images.   }
    \label{fig:fig5}
\end{figure}

Identifying the GAN-generated images was possible due to characteristic visual artifacts (Figure \ref{fig:fig5}). For the eye fundus images, we found that the synthetic image sometimes exhibited two optical discs, while every real fundoscopy always only exhibits one optical disc. No such occurrences were noted for the Medfusion-generated images. 
The GAN-generated images exhibited an artificial grid pattern for some generated histological images. Again, we did not observe such artifacts for the Medfusion model.
Chest radiographs were identifiable as synthetic by blurred letters or fuzzy borders and irregular appearances of medical devices (e.g., cardiac pacemakers). We found these artifacts to appear in both the GAN-generated and DDPM-generated synthetic images.
It should be noted that some of the real images showed strong visual artifacts due to image acquisition errors. However, the real artifacts differed from the artifacts in the synthetic data. We provide examples for such real artifacts in the supplementary material.

\section{Discussion}
The success of Deep Learning depends largely on the size and quality of training data. Therefore, generative models have been proposed as a solution to extend the availability of training data \cite{han2020c}. DDPM have been demonstrated to achieve superior image quality on natural images. In this study, we investigated if such models can also generate more diverse and realistic images as compared to GAN-based models in the medical domain.
We explored DDPM in three domains of medical data: ophthalmologic data (fundoscopic images), radiological data (chest x-rays) and histological data (whole slide images of stained tissue). We optimized our Medfusion architecture for medical image synthesis and found that image quality of DDPM generated images was superior to that of GAN generated images: Medfusion achieved an FID score of 11.63 in the eye, 30.03 in the histology, and 17.28 in the chest dataset which were all lower (better) than those of the GAN models (20.43, 49.26, 84.31), indicating a higher image quality. Also, the precision of the images generated by Medfusion was higher in the eye (0.70 versus 0.43), histology (0.66 versus 0.64), and chest (0.68 versus 0.30) dataset, indicating higher fidelity. A known problem with GANs is mode collapse, where the generator produces very realistic (high precision) but similar images so that the true diversity between the real images is not represented. Recall, as a measure of diversity, was strikingly low for histological images generated by the cGAN compared to Medfusion (0.02 versus 0.41), which indicates a mode collapse. 

In a study by Pinaya et al. \cite{pinaya2022}, a latent DDPM was trained to generate 3D MRI brain images and compared with two GANs. In agreement with our study, the latent DDPM model showed a lower (better) FID score of 0.008 compared to the two GANs (0.023 and 0.1576). Remarkably, FIDs were 3 to 4 orders of magnitude lower than in our study. We suspect that this is due to the 3D data used instead of 2D data because our measured FIDs are in the same order of magnitude as in previous studies on natural 2D images. Regardless of whether a GAN or our latent DDPM was used, we observed a maximum recall (diversity) of approximately 0.4 on the medical datasets. On natural images, recalls of 0.5 or better were observable \cite{dhariwal2021}. One possible reason for this is that natural images can achieve diversity by changing backgrounds and colors, medical images often have a constant (black or white) background, and colors are narrowly limited to e.g. grayscale. Therefore, diversity in medical images mainly manifests as change in details (eg. variations in heart size, or variations in the opacity of lung tissue). It may be more difficult to achieve high diversity while maintaining high fidelity in medical image generation than in natural images. Future studies are needed to investigate this.

Our study has limitations. First, the training and generation of the CheXpert and AIROGS images were performed in a lower resolution than the original resolution and the images were square (i.e. height equals width). There were two main reasons for this: 1) we wanted to compare the Medfusion model with the GAN results from the previous studies, which were trained and evaluated for a specific (lower) resolution. 2) the StyleGAN-3 that we employed for comparison only allows a quadratic resolution which must be a power of 2. Future studies should investigate how the Medfusion model behaves for higher resolutions compared to GAN models. 
Second, we would like to point out that the metrics used in this study to judge the image quality were not developed for medical images which could reduce their validity and should in general be evaluated with care. The development of metrics that are proxies for human judgment is still an ongoing topic area of research \cite{borji2021}.  Furthermore, to the best of our knowledge, no study has investigated these metrics focusing on medical images. This should be addressed in a future study.

\section{Conclusion}
Our study shows DDPMs provide promising new ways for medical image generation besides GANs. Future work should focus on high-resolution and quality metrics for 2D and 3D medical images.  

\section*{Funding}
The authors state that this work has not received any funding.

\section*{Conflicts of Interest}
The authors declare that they have no competing interests.

\newpage
\bibliographystyle{unsrtnat}
\bibliography{references}  






\newpage
\section*{Supplemental Material}

\subsection*{Artefacts in Real Data}
\begin{figure}[h!]
    \centering
    \includegraphics[width=\textwidth]{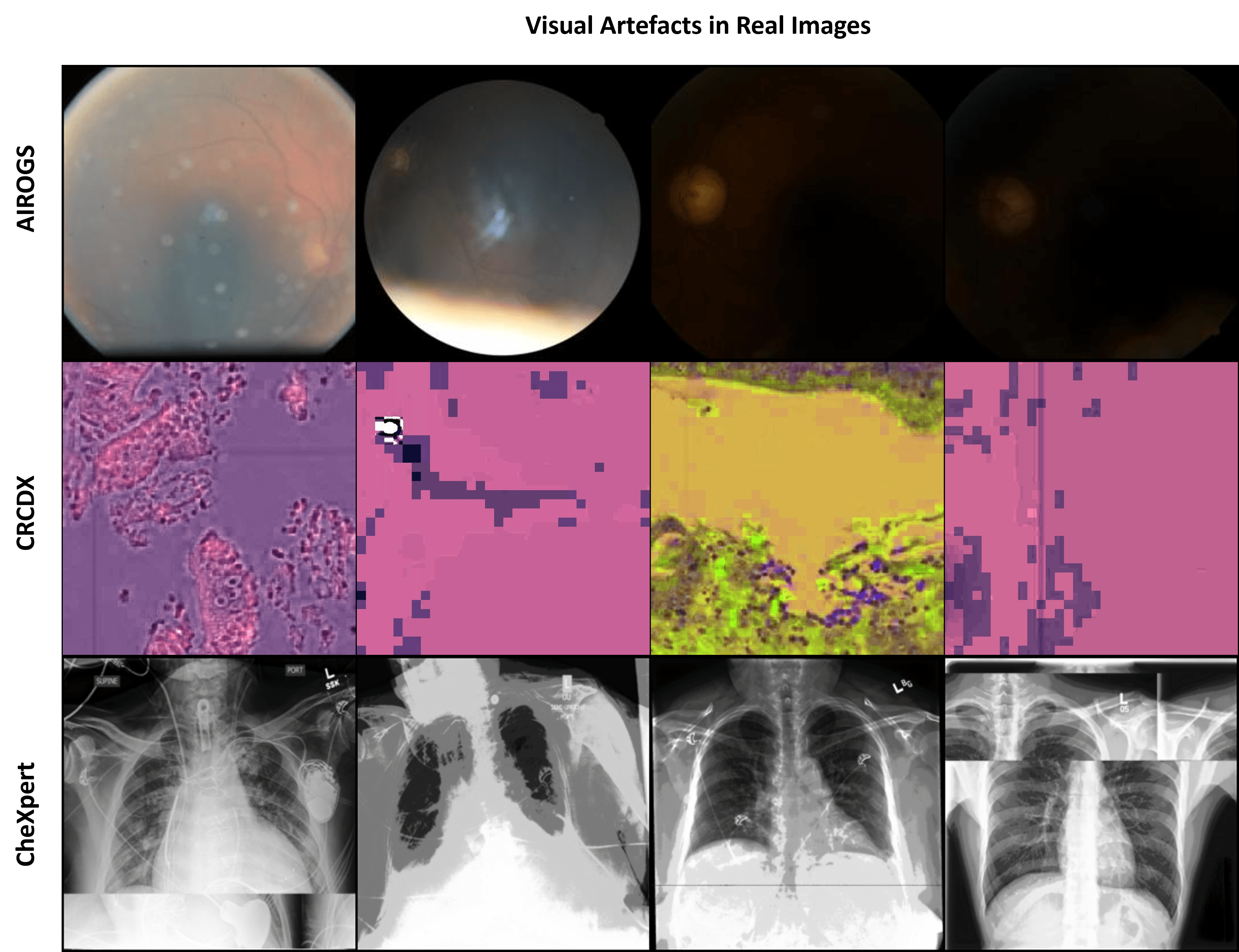}
    \caption{Examples of visual artifacts in the ARIOGS, CRCDX, and CheXpert datasets.   }
    \label{fig:fig6}
\end{figure}

Sifting through about 300 images of the AIROGS, CheXpert, and CRCDX data set, several images could be identified that showed strong visual artifacts (Figure \ref{fig:fig6}). Types of artifacts were limited to those that are severe and make the images appear "unnatural" because they are rare in the clinical routine or are even specific to the three data sets.

\subsection*{Autoencoder Loss}
The autoencoder loss consisted of an embedding loss and a reconstruction loss. 
We employed the Kullback–Leibler divergence between the embedding space and white Gaussian noise as embedding loss. The reconstruction loss was calculated between the input image and reconstruction output image as the summation of absolute distance (L1), Learned Perceptual Image Patch Similarity (LPIPS) \cite{zhang2018a}, structural similarity index measure (SSIM) \cite{wang2004}, and PatchGAN discriminator \cite{isola2017}. 
Besides the final, full-resolution prediction, the first-lower resolution of the decoder was also supervised by the reconstruction loss. Note that by applying SSIM to multiple resolutions, it becomes MS-SSIM \cite{wang2003} and the multi-resolution PatchGAN resembles the multi-resolution loss of Pix2PixHD \cite{wang2018b}. 

\subsection*{Autoencoder Testset }
The variational autoencoders of Stable Diffusion and Medfusion were evaluated on a test set independent of the training set (Table \ref{tab:table4}). The CheXpert test set included 200 frontal chest X-rays (134 without, 66 with cardiomegaly) and the CRCDX test set included 32,361 histology images (5,223 microsatellite unstable, 27,138 stable). A public test set for the AIROGS dataset did not exist.

\begin{table}[h!]
    \caption{Autoencoder reconstruction quality. Values represent mean±standard deviation, MSE = Mean Squared Error, MS-SSIM = Multiscale Structural Similarity Index Measure}
    \centering
    \begin{tabular}{@{}lcccc@{}}
    \toprule
    \multirow{2}{*}{}                & \multicolumn{2}{c}{CRCDX (Testset)} & \multicolumn{2}{c}{CheXpert (Testset)} \\
                                     & Stable Diffusion    & Medfusion     & Stable Diffusion     & Medfusion       \\ \midrule
    \multicolumn{1}{c}{MS-SSIM ↑}    & 0.876±0.045         & 0.903±0.037   & 0.974±0.006          & 0.995±0.001     \\
    \multicolumn{1}{c}{MSE ($10^{-5}$) ↓} & 617±404             & 528±316       & 89±28                & 24±8            \\ \bottomrule      
    \end{tabular}
    \label{tab:table4}
\end{table}

\subsection*{DDIM Sampling Steps}
\begin{figure}[h!]
    \centering
    \includegraphics[width=\textwidth]{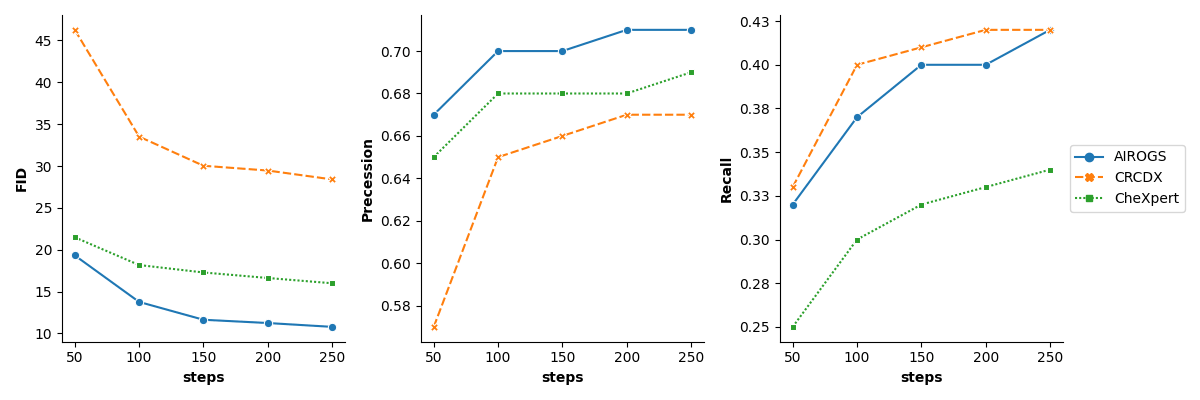}
    \caption{Fréchet Inception Distance (FID), Precision, and Recall as a function of the sampling steps.     }
    \label{fig:fig7}
\end{figure}

We increased the number of sampling steps of Medfusion's DDIM from t=50 to 250 in the inference mode and measured FID, precision, and recall on the reference data set (Figure \ref{fig:fig7}). In terms of the three metrics, there was an increase in image quality with increasing number of steps. In general, quality increased notably in the first 150 steps and then reached a plateau. Therefore, 150 steps appeared to be an appropriate tradeoff between globally increasing quality and increasing inference time.    

\end{document}